
\magnification=\magstep1

\hoffset=.4truecm
\voffset= 1truecm
\hsize=15truecm
\vsize=24truecm
\baselineskip=12pt
\hfuzz=18pt
\parindent=1truecm
\parskip=0.2truecm
\parskip=0.2truecm
\def\sqr#1#2{{\vcenter{\vbox{\hrule height.#2pt
        \hbox{\vrule width.#2pt height#1pt \kern#1pt
           \vrule width.#2pt}
        \hrule height.#2pt}}}}

\nopagenumbers
\overfullrule=0pt
\centerline{\bf $SO(5)_{q}$ and Contraction}\footnote{}{Research
supported in part by the EEC contract SCI-CT 92-0792 and CHRX-CT93-0340}

\vskip 1truecm
\centerline{A. Chakrabarti}

\centerline{\it Centre de Physique Th{\'e}orique,
Ecole Polytechnique}
\centerline{\it 91128 Palaiseau Cedex, France}
\centerline{\it Laboratoire Propre du CNRS UPR A.0014}%
\vskip 1truecm

\noindent
{\bf Abstract}

Representations of $SO(5)_{q}$ are constructed explicitly on the
Chevalley basis for all $q$, generic and root of unity. Matrix
elements of the generators are obtained for all representations
depending on three variable indices, the maximal number being 4. A
prescription for contraction is given such that a complete Hopf
algebra is immediately obtained for the non-semisimple contracted
case. For $q$ a	 root of unity the periodic representations for
$SO(5)_{q}$ and the contracted algebra are obtained directly in the
"fractional part" formalism which unifies the treatments for the
generic and root of unity cases. The $q$-deformed quadratic Casimir
operator is explicitly evaluated for the representations presented.

\eject

In our notations and conventions with the Cartan matrix

$$\Bigl\vert \matrix{
 2 &-2 \cr
-1 &2 \cr}
\Bigr\vert \eqno (1)
$$

and $[x] \equiv (q^{x} - q^{-x})/(q - q^{-1}), [x]_{2} \equiv (q^{2x}
- q^{-2x})/(q^{2}-q^{-2})$ the $q$-deformed $SO(5)$ algebra is (with
mutually commuting $q^{\pm h_{1}}$ , $q^{\pm h_{2}}$)

$$
\eqalign{
&q^{h_{1}} e_{1} = q e_{1} q^{h_{1}} ,  q^{h_{1}} f_{1} =
q^{-1} f_{1} q^{h_{1}},\cr
&q^{2h_{2}} e_{1} = q^{-1} e_{1} q^{2h_{2}} , q^{2h_{2}} f_{1} = q
f_{1} q^{2h_{2}},\cr
& q^{h_{1}} e_{2} = q^{-1} e_{2} q^{h_{1}} , q^{h_{1}} f_{2} = q
f_{2} q^{h_{1}},\cr
& q^{h_{2}} e_{2} = q e_{2} q^{h_{2}} , q^{h_{2}} f_{2} = q^{-1}
f_{2} q^{h_{2}},\cr
& [e_{1} , f_{1}] = [2h_{1}] , [e_{1} , f_{2}] = 0,\cr
& [e_{2},f_{2}] = [2h_{2}]_{2} , [e_{2} , f_{1}] = 0,\cr
& e_{1}^{3}e_{2} - (q^{2} + q^{-2} + 1) e_{1}^{2}e_{2}e_{1} + (q^{2} +
q^{-2} + 1) e_{1}e_{2}e_{1}^{2} - e_{2}e_{1}^{3} = 0\cr
& f_{1}^{3}f_{2} - (q^{2} + q^{-2} + 1) f_{1}^{2}f_{2}f_{1} + (q^{2}
+ q^{-2} + 1) f_{1}f_{2}f_{1}^{2} - f_{2}f_{1}^{3} = 0\cr
& e_{2}^{2}e_{1} - (q^{2} + q^{-2}) e_{2}e_{1}e_{2} + e_{1}e_{2}^{2}
= 0\cr
& f_{2}^{2}f_{1} - (q^{2} + q^{-2}) f_{2}f_{1}f_{2} + f_{1}f_{2}^{2}
= 0\cr}\eqno(2)
$$

We define $K$ and $M$ through

$$
q^{M} = q^{h_{1}} , q^{K - M} = q^{h_{2}} \eqno (3)
$$

The classical solutions are characterized by two invariant
parameters $n_{1}, \hfill\break n_{2} (n_{1} \geq n_{2})$ both being
integers or half-integers. Four variable indices span the most general
irreducible spaces. It can be shown that for the limiting $n_{2}$
values (for all $n_{1}$)

$$
(i) \quad n_{2} = 0, \qquad (ii) \quad n_{2} = {1 \over 2} \qquad and
\qquad (iii) \quad n_{2} = n_{1} \eqno (4)
$$

all Chevalley basis representations can be labelled with {\bf three}
variable indices. We will construct them in a form directly valid
for {\bf all $q$}, the irreducible spaces being spanned by the
states $\vert jmk>$.

The ansatz is

$$
\eqalign{
q^{M} \vert jmk> &= q^{m} \vert jmk> \cr
q^{K} \vert jmk> &= q^{k} \vert jmk> \cr
e_{1} \vert jmk> &= ([j - m] [j+m+1])^{1/2} \vert jm+1k> \cr
e_{2} \vert jmk> &= ([j-m+1] [j-m+2])^{1/2} a(j,k) \vert j+1 m-1 k+1> \cr
&+ ([j+m] [j+m-1])^{1/2} b(j,k) \vert j-1 m-1 k+1> \cr
&+ ([j+m] [j-m+1])^{1/2} c(j,k) \vert j m-1 k+1> \cr
}\eqno (5)$$

and

$$
<j'm'k' \vert f_{i} \vert jmk> = <jmk \vert e_{i} \vert j'm'k'> \eqno (6)
$$

For the general case one must have four-index states, say, $\vert
jmk\ell>$ and the ansatz (for $e_{2},f_{2}$) has to be generalized for the
variation of $\ell$. Here the reduced elements can be obtained from

$$
[e_{2},f_{2}] = [2h_{2}]_{2} = [K - M]_{2} \eqno (7)
$$
and
$$
e_{2}^{2}e_{1} - (q^{2} + q^{-2}) e_{2}e_{1}e_{2} + e_{1}e_{2}^{2} = 0
\eqno (8)
$$

All the other relations in (2) are then automatically satisfied. For the
three cases (4) one has respectively the following solutions

(i) $n_{2} = 0, n_{1} = n$

$$
\eqalign{
&a(j,k) = (q + q^{-1})^{-1} \Bigl({[n - j - k] [n + j + k + 3] \over [2j +
1] [2j + 3]}\Bigr)^{1/2} \cr
&b(j,k) = (q + q^{-1})^{-1} \Bigl({[n + j - k + 1] [n - j + k + 2] \over
[2j - 1]
[2j + 1]}\Bigr)^{1/2} \cr
&c(j,k) = 0 \cr
}\eqno (9)$$

For {\bf generic} $q$, the domains of parameters are

$$
\eqalign{
&j = 0, 1, 2, \cdots, n \cr
&k = n-j, n-j-2, \cdots, -(n-j-2), -(n-j) \cr
&m = j,j-1,\cdots, -(j-1),-j .\cr
}$$

\noindent
{\bf The parameters for the root of unity $q$'s will be discussed
separately}.

(ii) $(n_{2} = {1 \over 2} , n_{1} = n - {1 \over 2})$
\noindent
Defining $\delta \equiv {1 \over 2} (1 - (-1)^{n-j-k})$

$$
\eqalign{
a(j,k) &= (q+q^{-1})^{-2} [j+1]_{2}^{-1}
([n-j-k-\delta][n+j+k+3-\delta])^{1/2}\cr
b(j,k) &= (q+q^{-1})^{-2} [j]_{2}^{-1}
([n-j+k+1+\delta][n+j-k+\delta])^{1/2}\cr
c(j,k) &= (q+q^{-1})^{-2} [j+1]_{2}^{-1} [j]_{2}^{-1}\cr
&([n+(2\delta-1)j-k+\delta] [n + (2\delta-1)
j+k+1+\delta])^{1/2}\cr}\eqno (10)$$

\noindent
with

$$
\eqalign{
&j = n - {1 \over 2}, n - {3 \over 2}, \cdots, {1 \over 2}\cr
&k = n-j, n-j-1, \cdots, -(n-j) \cr
&m = j,j-1,\cdots,-(j-1),-j \cr
}$$

(iii) $n_{2} = n_{1} = n$

$$\eqalign{
&a(j,k) = (q+q^{-1})^{-1} \Bigl({[n-j]_{2} [n+j+2]_{2} [j+k+1] [j+k+2]
\over [2j+3] [2j+1] [j+1]_{2}^{2}}\Bigl)^{1/2} \cr
&b(j,k) = (q+q^{-1})^{-1} \Bigl({[n-j+1]_{2} [n+j+1]_{2} [j-k] [j-k-1]
\over [2j+1] [2j-1] [j]_{2}^{2}}\Bigl)^{1/2} \cr
&c(j,k) = (q+q^{-1})^{-1} [n+1]_{2} {([j-k][j+k+1])^{1/2} \over [j+1]_{2}
[j]_{2}}\cr
} \eqno (11)
$$

with

$$\eqalign{
&j = n,n-1,\cdots,0(1/2) \cr
&k = j,j-1,\cdots,-(j-1),-j \cr
&m = j,j-1,\cdots,-(j-1),-j \cr
}$$.

Defining

$$\eqalign{
&e_{3}^{(\pm)} = (q^{\pm 1} e_{1}e_{2} - q^{\mp1} e_{2}e_{1})\cr
&f_{3}^{\pm} = (q^{\pm 1} f_{2}f_{1} - q^{\mp1} f_{1}f_{2})\cr
&e_{4} = (q^{-1}e_{1}e_{3}^{(+)}-qe_{3}^{(+)}e_{1}) =
(qe_{1}e_{3}^{(-)}-q^{-1}e_{3}^{(-)}e_{1})\cr
&f_{4} = (q^{-1}f_{3}^{(+)}f_{1}-qf_{1}f_{3}^{(+)}) = (qf_{3}^{(-)}
f_{1}-q^{-1}f_{1}f_{3}^{(-)})\cr
}\eqno(12)
$$

the $q$-deformed second order Casimir operator can be written as

$$\eqalign{
A &= (q+q^{-1})^{-1} \Bigl\lbrace (f_{1}e_{1}+[M][M+1]
{(q^{(2K+3)}+q^{-(2K+3)}) \over (q+q^{-1})} + [K+1][K+2]\Bigr\rbrace \cr
&+f_{2}e_{2}+(q+q^{-1})^{-2} \Bigl\lbrace qf_{3}^{(+)} e_{3}^{(+)} q^{2M} +
q^{-1} f_{3}^{(-)} e_{3}^{(-)} q^{-2M} + f_{4}e_{4}\Bigr\rbrace - 1 \cr
} \eqno(13)
$$

2A can be shown to be the $q$-deformation of the classical Casimir

$$
J_{2} = \sum_{i<j} J_{ij}^{2} \eqno(14)
$$

(the $J_{ij}$'s being the standard Cartan-Weyl generators of $SO(5)$) with
the eigenvalue, for a irrep. $(n_{1},n_{2})$,

$$
n_{1}(n_{1}+3) + n_{2}(n_{2}+1). \eqno(15)
$$

For the cases (9), (10) and (11) respectively one obtains

(i) $$A\vert jmk> = {[n][n+3] \over (q+q^{-1})} \vert jmk> \eqno(16)$$

(ii) $$A\vert jmk> = {1 \over (q+q^{-1}} \Bigg\lbrace \Bigl( {q^{2}+q^{-2}
\over q+q^{-1}} \Bigr) [n_{1}] [n_{1}+3] + (q^{2}+q^{-2}-1) [{1 \over
2}] [{3 \over 2}] \Bigg\rbrace \vert jmk> \eqno(17)$$

(iii) $$A\vert jmk> = [n]_{2} [n+2]_{2} \vert jmk> \eqno(18)$$

$q = e^{i2\pi/N} \quad (N = 3,4,\cdots)$

{\bf For $q$ a root of unity the matrix elements remain formally the same
as before. Only the domains of the parameters involved are changed as
follows}.

Instead of being ordered integers or half integers as before they now have
"fractional parts" (in fact real proper fractions and also possibly
imaginary parts) with certain restrictions.

For brevily, I only indicate that for {\bf odd $N$} and {\bf periodic}
representations one should have
($\epsilon_{1},\epsilon_{2},\epsilon_{3},\epsilon_{4}$ each being $\pm 1$)

$$
\lbrace j, (j + \epsilon_{1}m) , (j+\epsilon_{2}k),
(n_{1}+\epsilon_{3}j),(n_{1}+\epsilon_{4}(j+\epsilon_{2}k))\rbrace \notin
Z/2 \eqno(19)
$$

Periodic, partially periodic and highest weight representations can
similarly be obtained for both odd and even $N$ with suitable
modifications of (19).

For {\bf periodic} (or cyclic) representations one can define the constans
$\gamma_{i}$ as

$$
\vert jmk> = \gamma_{1}^{-1} \vert j+N,m,k> = \gamma_{2}^{-1}
\vert j,m+N,k> = \gamma_{3}^{-1} \vert j,m,k+N> \eqno(20)
$$

For a general 4-index representation one also has the supplementary
periodicity condition

$$
\vert jmk\ell> = \gamma_{4}^{-1} \vert j,m,k,\ell+N> \eqno(21)
$$

Then one has {\bf 10 invariant parameters}, namely (with f.p. $\approx$
fractional parts)

$$
n_{1},n_{2} ; f.p.'s of j,m,k and \ell ;
\gamma_{1},\gamma_{2},\gamma_{3},\gamma_{4} \eqno(22)
$$

The dimension of these representations, is $N^{4}$.

For the restricted classes presented above (without the index $\ell$) one
obtains 8 parameters for the periodic case. The dimension is now $N^{3}$.

It can be shown that all the representaitons obtained above, for generic
$q$ and root of unity, are irreducible ones.

{\bf Contraction}

 From the set of generators

$$
(q^{\pm h_{1}}, q^{\pm h_{2}}, e_{1}, f_{1}, e_{2}, f_{2}) \eqno(23)
$$

we go over to the "contracted" set

$$
(q^{\pm h_{1}}, q^{\pm h_{2}}, e_{1}, f_{1}, \hat e_{2}, \hat f_{2})
\eqno(24)
$$

such that instead of

$$\eqalignno{
[e_{2},f_{2}] &= [2h_{2}]_{2} = [K - M]_{2}, \cr
{\rm now}\hskip 2truecm [\hat e_{2},\hat f_{2}] &= 0&(25) \cr}
$$

All other relations in the set (2) remain unchanged.

In each of the foregoing representations, let (with $\mu$ an arbitrary
constant, real for generic $q$)

$$
(\hat a, \hat b, \hat c) = \mu \Bigg( \lim_{n_{1} \rightarrow \infty}
q^{\mp n_{1}} (a,b,c)\Bigg ) \ {\rm for} \ q > 1\ {\rm and}\  q < 1\ {\rm
respectively}  \eqno(26)
$$

Then the contracted representations are given by only the following
modification of (5) and (6)

$$\eqalign{
\hat e_{2} \vert jmk> &= ([j-m+1] [j-m+2])^{1/2} \hat a(j,k) \vert j+1,
m-1, k+1> \cr
&+ ([j+m] [j+m-1])^{1/2} \hat b(j,k) \vert j-1, m-1, k+1> \cr
&+ ([j+m] [j-m+1])^{1/2} \hat c(j,k) \vert j,m-1, k+1> \cr
}\eqno(27)
$$

Thus corresponding to (11) one has

$$\eqalign{
&\hat a(j,k) = {\mu \over [j+1]_{2}} \Bigg({[j+k+1][j+k+2] \over
[2j+3][2j+1]} \Bigg)^{1/2} \cr
&\hat b(j,k) = {\mu \over [j]_{2}} \Bigg({[j-k][j-k-2] \over
[2j+1][2j-1]} \Bigg)^{1/2} \cr
&\hat c(j,k) = \mu \Bigg({[j-k][j+k+1] \over
[j]_{2}[j+1]_{2}} \Bigg)^{1/2} \cr
}\eqno(28)
$$

As compared to (13) the Casimir is now

$$
\hat A = \hat f_{2} \hat e_{2} + (q+q^{-1})^{-2} \lbrace \hat f_{3}^{(+)}
\hat e_{3}^{(+)} q^{2M+1} + \hat f_{3}^{(-)} \hat e_{3}^{(-)} q^{-2M-1}+
\hat f_{4} \hat e_{4} \rbrace \eqno(29)
$$

where, analogously to(12),

$$
\hat e_{3}^{(\pm)} = (q^{\pm 1} e_{1} \hat e_{2} - q^{\mp 1}\hat e_{2}
e_{1}) \eqno(30)
$$

and so on.

Instead of (18) one now has (indicating the new parameter $\mu$ in the
states)

$$
\hat A \vert jmk>_{(\mu)} = (q+q^{-1})^{2} \mu^{2} \vert jmk>_{(\mu)}
\eqno(31)
$$

Contractions of (9) and (10) are also easy to write down. They exhibit
specific degeneracies.

{\bf It can be shown that the Drinfeld-Jimbo Hopf algebra (coproducts,
counits and antipodes) of $SO(5)_{q}$ can be carried over intact for the
contracted, non-semisimple case}. Thus, for example

$$
\Delta \hat e_{2} = \hat e_{2} \otimes q^{K-M} + q^{-K+M} \otimes \hat
e_{2}\eqno(31)
$$

and so on.

{\bf For $q$ a root of unity, exactly as for the uncontracted $SO(5)_{q}$
case, one can obtain periodic and other types of representations by
suitably introducing fractional parts.}

For generic $q$ the contracted representations are, of course, all
infinite dimensional. For $q$ a root of unity one can obtain finite
dimensional representations. Thus, for example, (27) can give $N^{3}$
dimensional periodic representations.

At the classical level $(q=1)$ the contracted algebra has the following
structure. By writing the standard Cartan-Weyl generators $(J_{ij})$ in
terms of the Chevalley generators the effect of our contraction on the
$J$'s can be easily evaluated. It can be shown that one now has (denoting
each $J_{ij}$ undergoing contraction by $\hat J_{ij}$)

(1) an $S0(3)$ algebra $$(J_{12},J_{23},J_{13}),\eqno (32)$$

(2) two "translation triplets" (all mutually commuting)

$$
(\hat J_{14},\hat J_{24},\hat J_{34}) {\rm and} (\hat J_{15},\hat
J_{25},\hat J_{35}) \eqno(33)
$$

(3) with $J_{45}$ linking these triplets as (with $i = 1,2,3$)

$$
[J_{45},\hat J_{i4}] = -i\hat J_{i5} \ , \ [J_{45},\hat J_{i5}] = i\hat
J_{i4} \eqno(34)
$$

For $q\not= 1$ the translation triplets become non-commutative. This
deformation can again be evaluated without difficulty by expressing the
$J$'s in terms of (24) and (30) with its analogues for $\hat e_{3}^{(\pm)},
\hat f_{3}^{(\pm)},\hat e_{4}$ and $\hat f_{4}$, providing an invertible
set of relations. Due to lack of space I will not write down the equations
explicitly. It can provide an interesting exercise for the interested
reader.

A much more detailed discussion of the preceding results can be found in
[1]. The necessity for constructing representations of $SO(n)$ on the
Chevalley basis, with all Cartan generators diagonalized is emphasized and
explained there. This only opens the way for a successful $q$-deformation
as exhibited here for the simplest non-trivial case of $SO(5)_{q}$. The
contraction scheme which assures the existence of a complete Hopf algebra
was introduced before for $SU(n)_{q}$. Here, a beginning is made for
$SO(n)_{q}$. Though the explicit representations here are given for
restricted classes, {\bf the passage to a root of unity $q$ and the
contraction scheme are both quite general and works for any $q$-deformed
algebra}. All these points are more fully discussed in [1].

\noindent
References

\item {1.}
A. Chakrabarti, $SO(5)_{q}$ and contraction : Chevalley Basis
Representations of $q$ Generic and Root of Unity (preprint 1993 ; to be
published in Jour. Math. Phys.)
\item{ }
Contractions valid for {\bf all $q$} and
preserving the full Hopf algebra was introduced in.

\item{2.}
A. Chakrabarti, Jour. Math. Phys. {\bf 32} (1991) 1227.
\item{ }
 A different type
of contraction with $q \rightarrow 1$ can be found, for example, in

\item{3.}
E. Celeghini, R. Giachetti, E. Sorace and M. Tarlini, Jour. Math. Phys.
{\bf 32} (1991) 1159.

\item{4.}
J. Lukiersky, A. Nowicki and H. Ruegg, Phys. Lett. {\bf B293} (1992) 344.

The doubling of generators (such as $e_{3}^{(\pm)}, f_{3}^{(\pm)}$ in (12))
was recognized as an "antipode-extended system" in

\item{5.}
J. Lukiersky, A. Nowicki and H. Ruegg, Phys. Lett. {\bf B271} (1991) 321.

The extended system appears, for $SU(N)_{q}$ already in [2].

The "fractional part" formalism unifying representations for generic and
root of unity $q$ was introduced in

\item{6.}
D. Arnaudon and A. Chakrabarti, Comm. Math. Phys. {\bf 139} (1991) 461.

This is slightly modified here and made quite generally applicable.

Different approaches to $SO(5)_{q}$ can be found in

\item{7.}
D. Arnaudon and A. Chakrabarti, Phys. Lett. {\bf B262} (1991) 68.

\item{8.}
W.A. Schnizer, Representations of quantum groups, RIMS-961.

More relevant sources can be found in [1].

 \bye